\documentstyle[12pt]{article}
\makeatletter
\oddsidemargin   - .1 in
\evensidemargin   - .1 in
\leftmargin - .1 in
\newcommand{\doublespacing}{\let\CS=\@currsize\renewcommand{\baselinestretch}{2.0}\tiny\CS}

\newcommand {\beq} {\begin{equation}}

\newcommand {\eeq} {\end{equation}}

\date{}
\begin{document}

\thispagestyle{empty}\setcounter{page}{1}
\vskip10pt
\centerline{\bf The Time Dependence of  Fundamental Constants }
\centerline{\bf and}
\centerline{\bf Planck Scale Physics}

\vskip20pt
\centerline{\footnotesize Frederick Rothwarf$^{1,2}$ and Sisir Roy$^{3,4}$}  
\centerline{\footnotesize$^{1}$ {\it Department of Physics, George Mason University, Fairfax, VA 22030  USA}} 
\centerline{\footnotesize$^{2}$ {\it Magnetics Consultants, 11722 Indian Ridge Road, Reston, VA 20191,USA}} 
\centerline{\footnotesize$^{3}$ {\it Center for Earth Observing and Space Research and}}
\centerline{\footnotesize{\it School of Computational Sciences, George Mason University, Fairfax, VA 22030 USA}}
\centerline{\footnotesize$^{4}$ {\it Physics and Applied Mathematics Unit, Indian Statistical Institute, Calcutta, INDIA}}
\vskip5pt
\centerline{\footnotesize $^{2}$ e.mail: frothw@ieee.org }
\centerline{\footnotesize $^{3}$ e.mail: sroy@scs.gmu.edu}
\vskip20pt
\doublespacing
\abstract{\noindent{\small{A real aether model of the vacuum  proposed by Allen Rothwarf$^{(1)}$ based upon 
a degenerate Fermion fluid, composed of polarizable particle-antiparticle pairs, leads
to a big bang model of the universe where the velocity of light varies
inversely with the square root of the cosmological time.  Here this model is
used to determine the time dependence of certain fundamental constants,
i.e., permittivity $\epsilon(t)$ and permeability $\mu(t)$ of free space:  the
Gravitational constant $G(t)$; and the Planck units: length $l_p(t)$, time $t_p(t)$, and mass, $ m_p$.
\vskip10pt
\noindent
 
\newpage

\section{{\bf Introduction}}
An aether model of the universe has been proposed by Allen
Rothwarf$^{(1)}$ based upon a degenerate Fermion fluid, composed of
polarizable particle-antiparticle pairs in a negative energy state, 
relative to the null state or true vacuum. He proposed that the Fermion
fluid was composed primarily of a degenerate electron-positron plasma.
The model provides insight into a large number of physical and
cosmological phenomena for which conventional theories have 
unsatisfactory or no answers. Among the various issues he treated: 
wave-particle duality; the nature of spin (a vortex in the aether); the
derivation of Hubble's law; electric fields (polarization of the
aether); Zitterbewegung (a bare particle orbiting within a vortex core);
inflation and the big bang in cosmology; the Pauli exclusion principle
(repulsion between parallel spin vortices); the nature of the photon (a
region of rotating polarized aether propagating with a screw like
motion) are  few of them. A key assumption of this model is
that the Fermi velocity, \ $ v_F$, of the degenerate electron-positron plasma
that dominates the aether is equal to the present speed of light $c_0$. One of 
many important consequences of this model is that the speed of light decreases
with cosmological time according to the relationship
\begin{equation}
 c(t) = c_0{\left ( \frac{t_0}{ t} \right )}^{1/2}  
 \end{equation}
\noindent
where $c_0$ is the present speed of light and $t_0$ is the present age of the
universe after the end of inflation. In this paper, this relationship is used
to determine the time dependence of certain fundamental
constants, namely, the permittivity \ $\epsilon(t)$ and  permeability \ $\mu(t)$ \ of free
space; the Gravitational constant \ $G(t)$; \ and the Planck length \ $l_p(t) \sim 10^{-35}$m, 
\ time \ $t_p(t) \sim 10^{-44}$ s, and mass, \ $m_p \sim 10^{-8}$kg. It should be noted that some 
other models assuming a polarizable aether or vacuum have been proposed.
These have been cited and discussed in detail by Rothwarf$^{(1)}$ and Puthoff$^{(2)}$. However, 
none of these models derive a relationship for $c(t)$. Furthermore, the cited literature
refers to two types of aether. We denote one by $ {\bf A_{EM}}$ and the other ${\bf A_{G}}$. Van 
Flandern himself$^{(3)}$ and together with Vigier$^{(4)}$ made a clear distinction
between these two type of aethers. The Rothwarf model deals with ${\bf A_{EM}}$, while the
polarizable vacuum, proposed by Puthoff$^{(2)}$, concerns ${\bf A_{G}}$. The possible interaction of 
these two interpenetrating fluids will be discussed in a subsequent paper$^{(5)}$.
\noindent
\vskip5pt
\section{\bf The Time Variation of Physical Constants}
\noindent
\subsection{\bf {\bf The Determination  of $\epsilon(t)$ and $\mu(t)$}}
\noindent
\vskip6pt
Recently Puthoff$^{(2)}$ has published a Polarizable-Vacuum (PV) approach
to General Relativity (GR) in which the basic postulate is that the
polarizability of the vacuum, in the vicinity of a mass, (or other
mass-energy concentrations) differs from its asymptotic far-field value
by virtue of vacuum polarization effects, induced by the presence of the
mass. Thus, he proposed that for the vacuum itself
\begin{equation}
  D = \epsilon E = K \epsilon_0 E
\end{equation}
where $K$ is the altered dielectric constant of the vacuum (assumed to be a
function of position in his formulation), 
 due to changes in vacuum polarizability (GR-induced). \\
\indent
In the present paper, $K$ is considered 
as a function of cosmological time, $K(t)$. We consider that the expected polarizability of the 
vacuum (aether) will be changing as the density of the aether decreases with the expansion of the
 universe.  This is consistent with the assumption of this model which states that the number 
of electron-positron pairs in the universe remains constant after the end of the inflationary phase 
of the big bang.

We begin our analysis by considering the fine structure constant,\ $\alpha$, \ that governs 
electromagnetic interactions, i.e.,
\begin{equation}
 \alpha = \frac{e^2} { 2{\epsilon_o}h{c_0}} 
\end{equation}
where, $c_0  = (\frac{1}{\mu_0 e_0})^{1/2}$ in the aether model, considered here. In the present case, 
$e$ and $h$ are taken as constants; \ $c_0$, \ $\epsilon_0$, \ and \ $\mu_0$ \ are the present values of the speed of light, 
the vacuum permittivity and the vacuum permeability, respectively. Taking into consideration that  $\epsilon_0$ 
is expected (with a time-varying aether polarizability) to change to $\epsilon(t) = K(t)\epsilon_0$, 
the fine structure constant $\alpha$ can be rewritten as,
\begin{equation}
  \alpha = \frac{e^2}{2  \epsilon(t) h c(t)}
\end{equation}
or
\begin{equation}
 \alpha = \frac{e^2}{2 K(t)\epsilon_0 h c(t)}
\end{equation}

There is reason to believe that $\alpha$ has not varied significantly with time
since the end of inflation. From the observations of quasar absorption
spectra , Webb et al.$^{(6)}$ showed that $\alpha$ was slightly lower in the
past, with $\frac{\Delta\alpha}{\alpha} = -0.72 \pm 0.18 \times 10^{-5}$ \ for $0.5 \ < \ z \ < \ 3.5$. 
Analyzing geological constraints, imposed on a natural nuclear
fission event at Oklo, Damour and Dyson$^{(7)}$ concluded that \ $\frac{\Delta\alpha}{\alpha}$ \ 
over the past $1.5$ billion years has been $~ <5.0\times10^{-17}$yr$^{-1}$.  Therefore, with substantial 
compatibility, one can substitute $(1)$ into $(5)$ to obtain $K(t)$ which is 
\begin{equation}
  K(t) = \left (\frac{t}{ t_0} \right )^{1/2}
\end{equation}
\noindent
Then the permittivity is given by
\begin{equation}
\epsilon(t) = K(t)\epsilon_0  = \left (\frac{ t}{ t_0} \right )^{1/2} \epsilon_0
\end{equation}
\noindent
Now rewriting the expression for c(t) as
$$ c(t) = c_0 \left (\frac{t_0}{t} \right )^{1/2} = \left [\frac{1}{\mu(t)\epsilon(t)} \right ]^{1/2}$$ 
\noindent
and $ c_0  =\left (\frac{1}{\mu_0 \epsilon_0} \right)^{1/2}$, \ one can solve for $\mu(t)$ from (7) and (6) and obtain
\begin{equation} 
\mu(t) = \mu_0\left (\frac{ t}{ t_0} \right )^{1/2} 
\end{equation}
\noindent
Thus, the Rothwarf aether model shows that $\epsilon(t)$ and $\mu(t)$ \ have the same
functional dependence on cosmological time.
\noindent
\vskip5pt
\subsection{\bf Time Dependence of Gravitational Constant}
\noindent
\vskip5pt      
 Puthoff$^{(8)}$ \ discussed gravity as a Zero-Point-Fluctuation (ZPF)
force. To be more accurate, he developed in detail the approach, originally put forth by 
Sakharov$^{(9)}$, who proposed a model in which gravity is not a separately existing force,
but rather an induced effect associated with ZPF of the vacuum, in
much the same way as the Van der Waals and Casimir forces. Sakharov$^{(9)}$
conjectured that the Lagrange function of the gravitational field is
generated by vacuum polarization effects due to fermions. Akama, et al.$^{(10)}$ 
developed this approach further and claimed that the generation of gravity is
due to a collective excitation of fermion-antifermion pairs. Then Puthoff established
 a quantitative, point particle-ZPF model and showed that
gravitational mass and its associated gravitational effects, i.e., the
inverse square law, can be derived in a fully self-consistent way from
electromagnetic-ZPF-induced particle motion (Zitterbewegung). He denoted 
these particles, undergoing ZPF, as partons. This  model relates the gravitational
constant, $G$, to the cut-off frequency, \ $\omega_c$ \  as the broad-spectrum ZPF
radiation fields, generated by the Zitterbewegung. To obtain
quantitative agreement with the present value of $G$,\ i.e., $G_0$, \ Puthoff's model
requires that $(1)$ \ the cut-off frequency for the ZPF background to be of the
order of the Planck frequency,\ $\omega_p$; \ $(2)$ \ the partons, undergoing ZPF, have
masses of the order of the Planck mass, i.e., $m_p$; \ and \ $(3)$ \ the vibrational
amplitude be of the order of the Planck length, \ $l_p$. \ Using the usual definations, we write

\begin{equation}
l_p = \left (\frac{hG}{c^3} \right )^{1/2}, \ \ \ \ t_p = \left (\frac{hG}{c^5} \right )^{1/2}, \ \ \ \ 
\& \ \ \ \ m_p = \left (\frac{hG}{c} \right )^{1/2}
\end{equation}
\noindent
According to Puthoff the equation for $G$ follows as,
\begin{equation}
 G = \frac{2\pi^2 c^5}{h\omega_c^2}
\end{equation}
\noindent                                                                         
It is worth mentioning that Rothwarf$^{(1)}$ discussed in detail the Zitterbewegung of an electron of 
mass, \ $m_e$ \ on the core of a vortex in the presence of an aether medium and showed that the $\omega_c$ \ 
calculated with the aether model, corresponds precisely with that obtained, taking into consideration 
the relativistic QM Dirac equation for a free particle which is
\begin{equation}
 \omega_c = \frac{2\pi m_e c^2}{h}
\end{equation}
\noindent
At this point, we assume that the cutoff-frequency of Puthoff's partons, will have the 
same functional relationship as the electrons will have in (11), and that $m_p$ \ 
will now replace $m_e$ \  to obtain \ $\omega_c$ \ for the partons, vibrating at the cores
of vortices in the aether $A_G$.  When (11) and $(1)$ are substituted in (10) one
obtains
\begin{equation}
G(t) = G_0 \left (\frac{t_0}{t} \right )^{1/2} \ \ \ {\rm  where} \ \ \ \ G_0 = \frac{\pi h c_0}{2 m_p}^2   
\end{equation}
\noindent
which indicates that $G(t)$ \ has the same dependence on time as does \ $c(t)$.
\noindent
\vskip5pt
\subsection{ \bf Time Dependence of Planck Units: \ $l_p$,\ $ t_p$ \ and \ $ m_p$}

The Planck units have been defined in (9), where their dependence
on $c(t)$ implies that they may also vary with time. Substituting (1) and (12) into the expression of Planck length, 
\ $l_p$ \ in (9), we get,
\begin{equation}
l_p(t) = {l_p}_0 \left (\frac{t}{t_0} \right )^{1/2} \ \ \ \ {\rm where} \ \ \ \ {l_p}_0 
= \left (\frac{h G_0}{c_0^3} \right )^{1/2}
\end{equation}
\noindent
showing that $l_p$ has a time dependence which is the inverse of $c(t)$. In the same way, we can obtain the values 
for $t_p$ \ as
\begin{equation}
t_p = {t_p}_0(\frac{t}{t_0}) \ \ \ \ {\rm where} \ \ \ \ {t_p}_0 =\left (\frac{h {G_0}}{c_0^5} \right )^{1/2}
\end{equation}
\noindent
This establishes the fact that \ $t_p$ \ increases linearly with time. Finally, when we consider the expression 
for the Planck mass, \ $m_p$ \ in (9), we find that it is just a constant and  independent of time, since the
time variations of \ $c(t)$ \ in the numerator and $G(t)$ \ in the denominator, cancel out each other.
\noindent
\vskip5pt
\section{\bf Discussion }
We have applied the Rothwarf model of the electron-positron aether, ${\bf A_{EM}}$, \ to evaluate
the time dependence of several important physical cosnatnts. One of the major consequences of
this work is that the fractional time rate of change of $c(t)$ and $G(t)$ is calculated to be  
\ $  -(\frac{1}{2 t_0}) = { -3.65} \times 10^{-11}{\rm yr}^{-1}$ \ by using $t_0 = 13.7 \times 10^9 {\rm yr}$ \ 
from Webb et al.$^{(6)}$. Rothwarf$^{(1)}$ \ has suggested that a modified {\bf LIGO} experiment
might well be able to test this result. The future experiments with sufficient sensitivity will
be useful to verify our various time dependent predictions as given in (1),(6),(8) and (12). It
should be noted, however, that $G_0$ as defined in terms of $m_p$ in (12) represents a tautology, since,
$m_p$ itself is defined in terms of $G_0$. Thus, $G_0$ is a measured rather than a derived 
quantity in this model. \\
\indent
As we mentioned in the {\it introduction}, several models of gravity$^{(8-10)}$ \ require a 
polarizable vacuum (aether), made up of fermions and their anti-matter counterparts.
These particles, called partons$^{(8)}$ or gravitons$^{(3,4)}$, are assumed to have a mass equal to the 
Planck mass $m_p$ and to constitute an aether, $\bf{A_G}$, that transmits gravitational forces
at a speed $c_{G}$, \ which exceeds the speed of light $c_0$. \ Van Flandern and Vigier$^{(4)}$ have
analyzed planetary and cosmological data to obtain a lower limit of $c_{G} < 2 \times 10^{10} c_0 
= 6 \times 10^{18}$m/s. \ $\bf{A_G}$ is assumed to obey the same Fermi-Dirac statistics that govern $\bf{A_{EM}}$, 
so that the same arguments which give $c(t)$ in (1) will yield the same functional dependence for $c_{G}(t)$
and will give the Zitterbewegung frequency form given in (11). It should be emphasized that the fermion-antifermion model 
used here to describe $\bf {A_{EM}}$ and $\bf{A_G}$ assume that Planck cosntant, the mass and charge of the electron 
and also the Planck mass do not vary with cosmological time. We are aware that other approaches have been constructed by 
several authors$^{(11,12)}$, which fix G and allow the Planck constant to vary with time. These are based on a conjecture of
Calogero$^{(13)}$, who calculates Planck constant by using the arguments of stochastic mechanics
and by assuming that $G$ is a constant. We believe that our present approach yields more
consistent cosmological consequences that leads us to explore the interaction of the two aethers. The details of such
an interaction will be given in a subsequent paper$^{(5)}$  which might shed new light in understanding the physics at 
the Planck scale.
\noindent
\vskip15pt
{\bf Acknowledgements}
One of the authors (S.Roy) greatly acknowledges School of computational 
Sciences and Center for Earth Observing and Space Research, George Mason University, USA for their 
kind hospitality and funding for this work.

\newpage

{\bf REFERENCES}
\begin{enumerate}

\item   A.Rothwarf , Physics Essays,{\bf 11},444-466(1998).
\item   H.E. Puthoff, Foundations of Physics,{\bf 32},927-943(2002).
\item   T.Van Flandern , Phys.Lett.,{\bf A 250},1-11 (1998).
\item   T.Van Flandern and J.P. Vigier, Found. Phys.,{\bf 32},1031-1068(2002).
\item   F.Rothwarf and S.Roy (in preparation).
\item   J.K. Webb, M.T. Murphy, V.V. Flambaum, V.A. Dzuba, J.D. Barrow, C.W.
        Churchill, J.X. Prochaska, and A.M. Wolfe, Phys. Rev. Lett.,{\bf 87},091301(2001).
\item   T. Damour and F. Dyson, Nucl. Phys.,{\bf B480},37(1996).
\item   H.E. Puthoff, Phys. Rev.,{\bf A39},2333(1989).
\item   A.D. Sakharov, Theor. Math. Phys.,{\bf 23},435(1975).
\item   K. Akama, Y. Chikashige, T. Matsuki, and H. Terazawa, Prog. Theor.
        Phys.,{\bf 60},868(1978).
\item   G.Gaeta, Int.J.Theort.Phys.,{\bf 39},1339(2000).
\item   I.I.Haranas , J.Theoretics,{\bf 4},1(2002).
\item   F.Calogero, Phys.Lett.,{\bf A 228},335-346(1997).

\end{enumerate}
\end{document}